\documentclass[preprint,12pt]{elsarticle}

\usepackage{amssymb}
\usepackage{amsmath}
\usepackage{amsthm}
\usepackage{colortbl}
\usepackage{url}
\usepackage[nolist]{acronym}
\usepackage{graphicx}

\biboptions{sort&compress}

\journal{Epidemics}

\newcommand{\etal}{\textit{et al.}}
\newcommand{\pf}{\textit{P.\ falciparum}}
\newcommand{\pv}{\textit{P.\ vivax}}
\newcommand{\pfv}{\textit{P.\ falciparum} and \textit{P.\ vivax}}

\usepackage{todonotes}

\makeatletter
\newcommand*{\addFileDependency}[1]{
  \typeout{(#1)}
  \@addtofilelist{#1}
  \IfFileExists{#1}{}{\typeout{No file #1.}}
}
\makeatother

\title{A model for malaria treatment evaluation in the presence of multiple species}

\author[label1]{C.R. Walker \corref{cor1}}
\ead{camelia.r.walker@unimelb.edu.au}

\author[label1,label2,label3]{R.I. Hickson}
\author[label1]{E. Chang}
\author[label4,label5]{P. Ngor}
\author[label4]{S. Sovannaroth}
\author[label6]{J.A. Simpson}
\author[label6,label7]{D.J. Price}
\author[label1,label6]{J.M. McCaw}
\author[label5,label8,label9]{R.N. Price}
\author[label1]{J.A. Flegg}
\author[label8,label6]{A. Devine}

\address[label1]{School of Mathematics and Statistics, University of Melbourne, Australia}
\address[label2]{Australian Institute of Tropical Health and Medicine, and College of Public Health, Medical \& Veterinary Sciences, James Cook University, Australia}
\address[label3]{Health and Biosecurity, CSIRO, Australia}
\address[label4]{Cambodian National Center for Parasitology, Entomology and Malaria Control, Cambodia}
\address[label5]{Mahidol-Oxford Tropical Medicine Research Unit, Faculty of Tropical Medicine, Mahidol University, Thailand}
\address[label6]{Centre for Epidemiology and Biostatistics, Melbourne School of Population and Global Health, University of Melbourne, Australia}
\address[label7]{Department of Infectious Diseases, University of Melbourne, at the Peter Doherty Institute for Infection and Immunity, Australia}
\address[label8]{Division of Global and Tropical Health, Menzies School of Health Research and Charles Darwin University, Australia}
\address[label9]{Centre for Tropical Medicine and Global Health, Nuffield Department of Clinical Medicine, University of Oxford, UK}

\begin{document}
\begin{abstract}
\textit{Plasmodium} \textit{falciparum} and \textit{P.\ vivax} are the two most common causes of malaria. While the majority of deaths and severe morbidity are due to \pf, \pv~poses a greater challenge to eliminating malaria outside of Africa due to its ability to form latent liver stage parasites (hypnozoites), which can cause relapsing episodes within an individual patient. In areas where \pfv~are co-endemic, individuals can carry parasites of both species simultaneously. These mixed infections complicate dynamics in several ways: treatment of mixed infections will simultaneously affect both species, \pf~can mask the detection of \pv, and it has been hypothesised that clearing \pf~may trigger a relapse of dormant \pv. When mixed infections are treated for only blood-stage parasites, patients are at risk of relapse infections due to \pv~hypnozoites.\\

We present a stochastic mathematical model that captures interactions between \pf~and \pv, and incorporates both standard schizonticidal treatment (which targets blood-stage parasites) and radical treatment (which additionally targets liver-stage parasites). We apply this model to assess the implications of mass drug administration (MDA), different treatment coverage of radical cure for mixed and \pv~infections and a ``unified radical cure" treatment strategy where \pf, \pv~and mixed infections all receive radical cure after screening \ac{gpd} normal. We find that a unified radical cure strategy leads to a substantially lower incidence of malaria cases and deaths overall. MDA with schizonticidal treatment was found to decrease \pf with little effect on \pv. We perform a univariate sensitivity analysis to highlight important model parameters.

\end{abstract}



\flushbottom
\maketitle
$^*$ Corresponding author: \texttt{camelia.walker@unimelb.edu.au}

\thispagestyle{empty}

\begin{acronym}[malaria]

\acro{plas}[\textit{P.}]{\textit{Plasmodium}}
\acro{pf}[\textit{P. falciparum}]{ \textit{Plasmodium falciparum}}
\acro{pv}[\textit{P. vivax}]{\textit{Plasmodium vivax}}
\acro{po}[\textit{Po}]{\ac{plas} \textit{ovale}}
\acro{plm}[\textit{Pm}]{\ac{plas} \textit{malariae}}
\acro{pk}[\textit{Pk}]{\ac{plas} \textit{knowlesi}}

\acro{act}[ACT]{artemisinin-based combination therapies}
\acro{cq}[CQ]{chloroquine}

\acro{anoph}[\textit{An.}]{\textit{Anopheles}}

\acro{ctmc}[CTMC]{continuous-time Markov chain}

\acro{fsat}[FSAT]{focal screening and treatment}

\acro{gpd}[G6PD]{glucose-6-phosphate dehydrogenase}
\acro{gms}[GMS]{Greater Mekong Subregion}

\acro{mda}[MDA]{mass drug administration}

\acro{ode}[ODE]{ordinary differential equation}

\acro{rdt}[RDT]{rapid diagnostic test}

\end{acronym}

\section{Introduction}
\label{sec:introduction}

Almost half of the world’s population is at risk of malaria, with ongoing transmission reported in 85 countries \cite{world_health_organization_world_2021}. In 2020 there were an estimated 241 million cases and 627,000 malaria deaths, with funding for control and elimination estimated at US\$3.3 billion \cite{world_health_organization_world_2021}. Over the last decade substantial gains have been made in reducing the burden of disease. In 2014 the leaders of 18 malaria endemic countries in the Asia Pacific committed to eliminating the disease in the region by 2030 \cite{gosling_malaria_2012}. In this region the two parasite species that cause the greatest burden of malaria are \textit{Plasmodium falciparum} and \pv. Most research and intervention efforts have been focused on \pf, the most pathogenic parasite species. However, outside of Africa \pv~ is becoming the predominant cause of malaria, and almost invariably co-exists with \pf. While malaria control measures impact both species, these are often less effective against \pv~ primarily due to the parasite's ability to form dormant liver parasites (hypnozoites) that can reactivate weeks to months after the initial infection, causing future infections (relapses). \pv~also forms sexual stages early in infection and is able to transmit to the mosquito vector before the patient seeks treatment. Furthermore, \pv's lower parasite density makes it more difficult to detect than \pf. 

Primaquine is the only widely-used drug available that kills hypnozoites. The combination of primaquine plus a schizonticidal drug, such as \ac{cq} or \ac{act}, is known as radical cure. Primaquine can cause drug-induced haemolysis in individuals with glucose-6-phosphate dehydrogenase (\ac{gpd}) deficiency, an inherited enzymopathy present in up to 30\% of malaria endemic populations.  For this reason the WHO recommends screening for \ac{gpd} deficiency prior to administration of primaquine to reduce the risk of severe primaquine-induced haemolysis \cite{who_guidelines_2022}. The effectiveness of primaquine is limited by the reluctance of healthcare providers to prescribe it, and patient adherence to complete a course of treatment \cite{thriemer_challenges_2017}. New point-of-care tools for diagnosing \ac{gpd} deficiency have recently come onto the market but have yet to be introduced widely into clinical practice. The challenges in safely and consistently treating \pv~with radical cure has resulted in its relative rise as a proportion of malaria cases \cite{Kenangalem_malaria_2019,price_plasmodium_2020,carrara_malaria_2013}. In one modelling study over 80\% of \pv~cases in the \ac{gms} was estimated to have arisen from relapses \cite{adekunle_modeling_2015}, highlighting the importance of radical cure to reduce the burden of disease \cite{douglas_chemotherapeutic_2012}.
 
 Successful malaria elimination campaigns in co-endemic settings will require widespread use of safe and effective radical cure to patients presenting with \pv~as well as the hidden reservoirs of infection. Failure to consider \pv~malaria as a target for elimination may compromise \pf~elimination campaigns because communities that continue to experience cases of malaria, even if due to another type of parasite, may show a reduced willingness to participate in future interventions designed to prevent re-introduction of \pf. In an effort to accelerate \pf~malaria elimination in the \ac{gms}, the use of \ac{mda} or mass screening and treatment is now being investigated \cite{landier_effect_2018}. These interventions do not include radical cure, but all stages (blood and liver) of all parasite species will need to be eradicated to eliminate malaria.

Cambodia aims to eliminate all species of malaria by 2025. In 2019, mixed infections of both \pfv~accounted for 16.6\% of malaria infections \cite{chhim_malaria_2021}. Mixed infections can change treatment outcomes in several ways: a mixed infection will be treated for both species simultaneously, \pf~malaria can mask a \pv~malaria co-infection \cite{abba_rapid_2014,ashton_performance_2010} and an episode of \pf~malaria is associated with a greater risk of \pv~infection in the subsequent weeks after treatment \cite{commons_risk_2019,Lin:2011,hossain_risk_2020}. It has been hypothesized that the fever and haemolysis caused by acute \pf~ malaria may trigger reactivation of \pv~hypnozoites and subsequent relapse.  Whereas current radical cure policy is reserved for patients presenting with \pv~malaria, a unified treatment policy, in which patients presenting with either \pv~or \pf~are prescribed radical cure, has potential to reduce recurrent episodes of malaria and target hidden reservoirs of infection \cite{poespoprodjo_supervised_2022}. 

While a range of mathematical models for malaria have been proposed, implemented, analysed and used to support policy decisions over the last 100 years---as reviewed recently \cite{mandal_mathematical_2011,smith_agent_2018}---few models have included the parasite dynamics of both \pfv~\cite{aguas_modeling_2012,pongsumpun_mathematical_2008,pongsumpun_impact_2010,silal_malaria_2019}. To our knowledge, only one of these modelling investigations explored interactions between species \cite{silal_malaria_2019}. Shretta \etal~developed a deterministic metapopulation model of \pfv, and incorporated key interactions between \pfv, including ``treatment entanglement'' (any treatment affecting the other parasite species), ``triggering'' (\pv~hypnozoite activation following an episode of \pf), and ``masking'' (where non-\pf~\ac{rdt} results are either missed or falsely attributed to be \pf). The remaining models~\cite{aguas_modeling_2012,pongsumpun_mathematical_2008, pongsumpun_impact_2010} explicitly and/or effectively consider the dynamics of the two species to be completely independent. 

We present the first stochastic agent-based model for the transmission of both \pfv, which addresses the dynamics of mixed infections, parasite interactions and antimalarial treatments. Our model considers humans as discrete agents which transition between compartments according to a \ac{ctmc} model. The \ac{ctmc} is coupled with a system of \acp{ode} that govern the mosquito population, where the transmission rate both from mosquitoes-to-humans and humans-to-mosquitoes are held constant over small time-steps. The stochasticity of our model allows it to appropriately capture infectious disease dynamics at low incidence, as malaria approaches elimination. Our model has 6 compartments for \pf~and 7 for \pv, representing a model with lower complexity than the other multi-species model with interactions, which has 14 and 17 compartments, respectively \cite{silal_malaria_2019}. The reduction in model complexity partially comes from removing age-stratification from the model. One of the main effects of age is in the acquisition of immunity to prevent developing clinical malaria, which we capture through lower probabilities of clinical malaria upon reinfection or relapse (i.e., in the following, the probability of clinical malaria upon reinfection or relapse is 0.5, compared to 0.95 for a naive \pf~infection). Even with the reduction in model complexity compared to \cite{silal_malaria_2019}, we note that our model requires many input parameters, many of which are poorly defined in the literature. Hence, we perform a univariate sensitivity analysis to understand the impact of changes to each parameter with respect to the model outputs, malaria cases and deaths.

As an example, we consider scenarios with Cambodia-like \pfv~prevalence and parameters, since both \pfv~are present and the \ac{anoph} populations are able to transmit both. Our model is applied to assess the effect of blood-stage treatment, differing coverage of radical cure prescription, and a unified treatment policy in which radical cure is prescribed to patients presenting with \pv, \pf~and mixed infections. For each of these treatment scenarios, we also consider an \ac{mda} intervention, where a proportion of the population are prescribed blood-stage treatment, which allows for asymptomatic infections to be treated.

\section{Methods} \label{sec:methods}

\subsection{Transmission model}

To capture the transmission dynamics of both \pfv, we use a stochastic agent-based approach for the human population coupled with a deterministic system of \acp{ode} for the mosquito population. Each human agent has their status with respect to both \pfv~tracked over time, which allows mixed infections to be captured. The agent-based model is implemented by holding rates constant over discrete time-steps for computational efficiency and for ease of coupling to the \acp{ode} that govern the mosquito population. 

\begin{figure}
\begin{centering}
\includegraphics[width=0.75\textwidth]{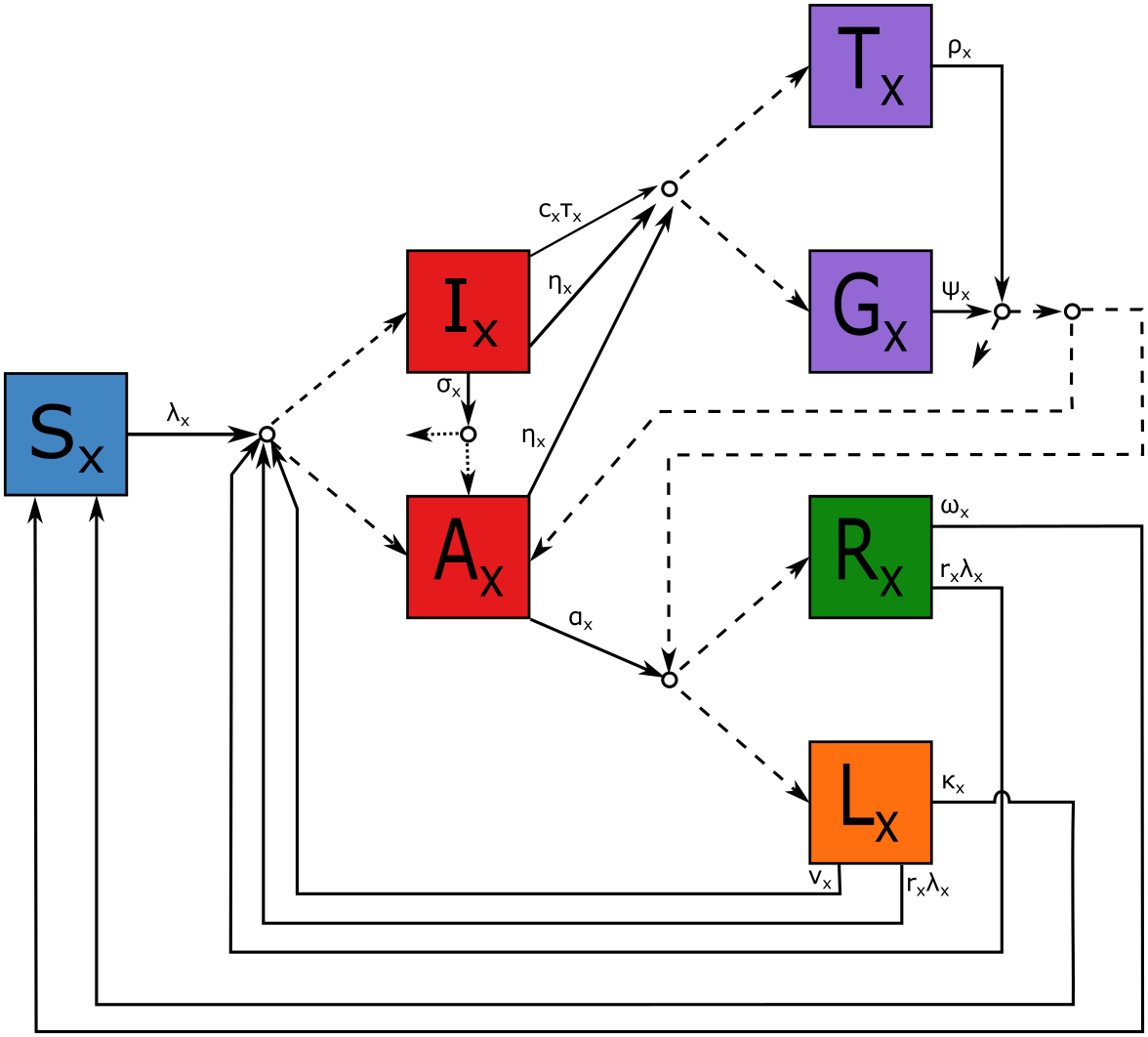}
\caption{Simplified schematic of the human transmission model for a single parasite species, $x$. The model compartments are $S$ (susceptible), $I$ (clinical infectious), $A$ (asymptomatic infectious), $T$ (undergoing blood-stage treatment), $G$ (undergoing radical cure), $R$ (recovered and partially-immune) and $L$ (latent stage hypnozoites for \pv~ only). Solid lines represent rates, the dashed lines probabilities, and the circles designate where a rate is split by probabilities. The probability parameters are not explicitly shown in the figure as, in many cases, the probability of each outcome depends on the current state (for example, the probability of symptoms upon infection is lower for recovered individuals than susceptible individuals).} 
\label{fig:single-species-model}
\end{centering}
\end{figure}

Each individual's state is bivariate to specify their state with respect to \pf~and to \pv. For each \textit{Plasmodium} species, humans are regarded as being susceptible ($S$), infectious with clinical symptoms ($I$), infectious but asymptomatic ($A$), recovered with no hypnozoites ($R$), recovered with hypnozoites ($L$ for latent: not applicable for \pf), undergoing blood-stage treatment with no radical cure ($T$), or undergoing treatment with radical cure ($G$). Radical cure is defined as low-dose primaquine (3.5mg/kg total) administered over 14 days. A simplified schematic of the human transitions are depicted in Figure~\ref{fig:single-species-model}. Given the large number of connections between states required to describe the transmission and treatment dynamics, the model schematic uses a single line between connectors where multiple transitions apply and does not depict interactions between the species. The full list of possible transition rates and stoichiometries is provided in the Supplementary Table S1.

The dynamics of a human individual infected with type $x$ malaria are briefly described here for $x=f$ (\pf) and $v$ (\pv).  Individuals susceptible ($S$) to type $x$ malaria are infected at rate $\lambda_x$. Upon infection they develop clinical symptoms ($I$) with probability $p_{c,x}$ or are otherwise asymptomatic ($A$). Individuals are symptomatic for a mean duration of $1/\sigma_x$, at which point they either become asymptomatic ($A$) or die without treatment with probability $p_{I,x}$. The individuals with clinical symptoms may be treated at rate $c_x \tau_x$, where $c_x$ is the probability that they are able to access healthcare and $\tau_x$ is the rate at which medical attention is sought if it is readily available. Individuals with asymptomatic malaria will clear all blood-stage parasites at rate $\alpha_x$ and, for \pv, will be left with hypnozoites with probability $p_{h,v}$. When an individual with \pv~seeks treatment they are prescribed radical cure ($G$) with probability $p_{N,x}$, otherwise they receive blood-stage treatment ($T$).  

Any infectious individual may additionally be treated at rate $\eta_x(t)$ via an intervention program (such as \ac{mda}): the form of $\eta_x(t)$ will be discussed in Section \ref{sec:future_scenarios}. When an individual is treated this way they are prescribed radical cure with probability $p_{M,x}$, otherwise they receive blood-stage treatment. An individual undergoes treatment for an average of $1/\psi$ days (14-day primaquine) at which point they may: die with probability $p_{G,x}$, remain with asymptomatic blood-stage malaria with probability $p_{TfP}$, if $x=v$ they are left with latent hypnozoites ($L$) with probability $p_{P,v}$, otherwise they recover ($R$). Similarly, an individual ends blood-stage treatment after an average of $1/\rho_x$ days (3-day ACT) at which point they may: die with probability $p_T$, remain with asymptomatic blood-stage malaria with probability $p_{TfA}$, if $x=v$ they are left with latent hypnozoites with probability $p_{A,v}$, otherwise they recover. Latent stage \pv~infected individuals experience a relapse at rate $\nu_v$ or they are reinfected at rate $\lambda_v r_v$ (where $r_x$ represents a possible reduction in susceptibility due to anti-parasite immunity) \cite{doolan_aquired_2009}. Upon relapse or reinfection from $L$ the individual experiences clinical malaria with probability $p_{L,v}$ (where $p_{L,v}<p_{c,v}$). Recovered individuals ($R$) are reinfected with rate $\lambda_x r_x$ at which point they become a clinical case with probability $p_{R,x}$ (where $p_{R,v}<p_{c,v}$). In addition to the possibility of relapse or reinfection, recovered and latent individuals lose immunity and hypnozoites at rates $\omega_x$ and $\kappa_v$, respectively, and return to being susceptible.\\

For individuals with mixed infections, there are several transitions in the model which depend on the individual's state with respect to both species: we refer to these dependencies as species ``interactions". When an individual with a mixed infection is treated, they move from states in $\{I, A, L\}$ to a state in $\{T, G\}$ for both species (depending on the treatment); this is referred to as ``treatment entanglement". Similarly, when a patient stops treatment with respect to one malaria species, they are moved to one of the post-treatment states with respect to the other species. Death with respect to one species will cause a transition to death with respect to the other. The model allows treatment efficacies to vary for mixed infections; however, in this work we have assumed antimalarial efficacy against each species to be equivalent to the efficacy against mono-infections. We model \pv~relapses triggered by the recovery of \pf~(``triggering") by setting the relapse rate for a person that is recovered ($R$) from \pf~ but has latent stage ($L$) \pv~to be $\hat{\nu}_{fv} = z_f\nu_v$, where $z_f>1$. The model also allows blood-stage \pv~to be masked by blood-stage \pf~(``masking") by treating a mixed infection as though it were a \pf~infection only with probability $h_v$. Explicitly, for an individual with \pfv~both in states $I$ or $A$, the probability of receiving radical cure is $p_{N,fv} = h_v p_{N,f} + (1-h_v)p_{N,v}$.\\

\subsection{Transmission intensity and vector species}
The dynamics of the mosquito population are governed by a system of \acp{ode} (presented in the Supplementary \S2). The mosquitoes follow standard Susceptible-Exposed-Infectious (SEI) dynamics with the addition of a seasonally varying death rate and the ability for mosquitoes to carry and spread mixed infection in a single bite (known as simultaneous inoculation). Asymptomatic individuals tend to have a lower peripheral parasitaemia and therefore were assumed to be less infectious to mosquitoes than individuals with clinical malaria with a relative infectiousness of $\zeta_{A,x}=0.1$.
 
We consider model parameters that would be representative of a Cambodia-like context, where both \pfv~circulate, and the mosquito species present (\ac{anoph} \textit{dirus}, \ac{anoph} \textit{minimus}, \ac{anoph} \textit{maculatus}, and \ac{anoph} \textit{barbirostris}) are able to transmit both parasite species~\cite{Siv_Plasmodium_2016} allowing mosquitoes to be modelled as a single population that spreads both \pfv.

\subsection{Treatment Scenarios}
\label{sec:future_scenarios}

We simulate three treatment scenarios: current practice, accelerated radical cure, and unified radical cure. In each scenario we assume that when a person tests positive for malaria the species is always identified correctly for mono-infections, since specialised \acp{rdt} have been shown to have high sensitivity and specificity, particularly for \pf~(see, for example, \cite{abba_rapid_2014}). The three treatment practices considered are:
\begin{enumerate}
\item \textbf{Current practice:} Under this scenario, \pf~and most \pv~cases are prescribed a blood-stage treatment but only 16\% of \pv~cases are prescribed radical cure. The low coverage of radical cure was chosen to match the rates reported from a study in Cambodia where radical cure was prescribed conservatively to 16\% of detected \pv~cases \cite{hoyer_focused_2012}. That is, the probability that an individual receives radical cure when being treated for malaria $x$, is
\[p_{N,x}= \begin{cases} 
      0, & \text{for } x=f \\
      0.16, & \text{for } x=v \\
     (1-h_v)0.16, & \text{for } x=fv,
   \end{cases}
\]
where $h_v$ is the probability that \pv~is masked by \pf~when the individual is tested. 

\item \textbf{Accelerated radical cure:} Under this scenario, any eligible person who is diagnosed with \pv~and returns a \ac{gpd} normal \ac{rdt} is prescribed radical cure with a low-dose 14-day course of primaquine (total dose 3.5 mg/kg) alongside a 3-day course of blood-stage treatment. Anyone who is $>$6 months old and is not pregnant or lactating is considered eligible for radical cure. We assume that the \ac{gpd} \acp{rdt} have a sensitivity of 94\% and a specificity of 91\% \cite{ley_performance_2019}, 6\% of the population have \ac{gpd} enzyme activity $<$30\%, 2\% of the population are pregnant and/or lactating and 2\% of the population are $<$6 months old. When combined, this means that 18\% of people diagnosed with \pv~will be ineligible to receive radical cure (further details are given Supplementary \S3). The probability of receiving radical cure under this scenario is
\[p_{N,x}= \begin{cases} 
      0, & \text{for } x=f \\
      0.82, & \text{for } x=v \\
     (1-h_v)0.82, & \text{for } x=fv.
   \end{cases}
\]

\item \textbf{Unified radical cure:} Under this scenario, radical cure is prescribed to any eligible person in whom peripheral parasitaemia is detected, with eligibility as defined and calculated in the accelerated radical cure scenario. The probability of receiving radical cure in this scenario is
\[p_{N,x}= \begin{cases} 
      0.82, & \text{for } x=f \\
      0.82, & \text{for } x=v \\
     0.82, & \text{for } x=fv.
   \end{cases}
\]
\end{enumerate}
The three treatment scenarios and the probability of receiving radical cure is summarised in Table \ref{tab:scenarios} with an assumed probability of masking of $h_v = 0.5$.

\begin{table}[ht]
\label{tab:flucan}
\begin{centering}
\hspace*{-0.5cm}
\begin{tabular}{|p{3cm}|p{3cm}|p{3cm}|p{3cm}|}%
\hline 
\cellcolor[gray]{0.9}  & \cellcolor[gray]{0.9} Current Practice  &  \cellcolor[gray]{0.9} Accelerated RC & \cellcolor[gray]{0.9} Unified RC \\ \hline 
\multicolumn{4}{|c|}{\cellcolor[gray]{0.95} Treatments} \\ \hline
\emph{P. falciparum} & ACT + PQ1 & ACT + PQ1 & ACT + PQ14 \\
\cellcolor[gray]{0.95} \emph{P. vivax} & \cellcolor[gray]{0.95}CQ + PQ14 & \cellcolor[gray]{0.95}CQ + PQ14  & \cellcolor[gray]{0.95}ACT + PQ14 \\ 
Mixed & ACT + PQ14 & ACT + PQ14 & ACT + PQ14 \\ \hline
\multicolumn{4}{|c|}{\cellcolor[gray]{0.95} Radical Cure Coverage (given detected infection)} \\ \hline
\emph{P. falciparum} & 0 &  0 & 0.82 \\
\cellcolor[gray]{0.95} \emph{P. vivax} & \cellcolor[gray]{0.95} 0.16 &  \cellcolor[gray]{0.95} 0.82 & \cellcolor[gray]{0.95} 0.82 \\
Mixed & 0.08 & 0.41 & 0.82 \\ \hline
\end{tabular} 
\caption{Treatments and radical cure coverage by species for each scenario, with an assumed probability of masking of $h_v=0.5$. Here, radical cure coverage is defined as the probability of receiving radical cure given a detected infection. Treatments ACT, CQ, PQ1 and PQ14 denote a 3-day course of artemisinin-based combination therapy, a 3-day course of chloroquine, a 1-day course of primaquine and a 14-day course of primaquine, respectively.}
\end{centering}
\label{tab:scenarios}
\end{table}

For each of the three treatment scenarios we also consider the impact of an \ac{mda} intervention where a proportion of the population are given a blood-stage treatment irrespective of infective status, thus allowing a proportion of clinical and asymptomatic blood-stage infections to be treated.
We assume that a proportion, $p$, of the population are prescribed a blood-stage treatment over a fixed period of time, $\Delta t = t_2 - t_1$, so that the treatment rate of an individual with species $x$ due to \ac{mda} is:
\[\eta_x(t)= \begin{cases} 
      \frac{-\text{ln}(1-p)}{\Delta t}, & t\in (t_1,t_2), \\
     0, & \text{otherwise.}
   \end{cases}
\]
We assume that people are not screened for \ac{gpd} status nor prescribed radical cure during \ac{mda}, based on concerns about haemolytic risks outweighing the benefits in patients who do not have malaria \cite{ong_systematic_2017}.  That is, the probability that an individual receives radical cure under \ac{mda}, given that they are treated for malaria type $x$ is $p_{M,x}$= 0 for all $x$. 

\subsection{Implementation}
We present the impact of different treatment and intervention strategies on the number of malaria cases and deaths from 2021 to 2030, where 2030 is the regional target for malaria elimination.  

For each treatment and intervention scenario we run 50 model simulations and record the model compartments over time.  Given the relatively short time frame,  we ignore background human demographic dynamics in our model to reduce computational complexity.

For the \emph{current practice} scenarios we assume that radical cure is prescribed conservatively and does not increase the risk of haemolysis (as a best case scenario). To account for heamolytic risk in the \emph{accelerated radical cure} and \emph{unified radical cure} scenarios we increase the probability of death by $5\times10^{-6}$ for patients administered primaquine and the probability of radical cure failure by $5\times 10^{-4}$ (details on these values are given in Supplementary \S3).

For the \ac{mda} intervention, we assume that half of the population receive blood-stage treatment over a 30 day period. The \ac{mda} is rolled out twice yearly, before and after the annual peak in transmission.

Initial conditions are set to resemble an endemic setting in eastern Cambodia, because the prevalence of clinical and asymptomatic infections in the region is well documented \cite{Sandfort:2020} and levels of immunity in eastern Cambodia were studied in a 2005 sero-survey \cite{cook_sero_2012}. Some parameter values were based on expert elicitation or a limited evidence base, and some parameter estimates vary greatly between studies. As such, the scenarios presented here are indicative of population dynamics of multi-species infections over time that accommodates interaction between \pf~and \pv~infections and expected trends in the impacts of different interventions. Accordingly, the scenarios here should not be interpreted as forecasts of malaria cases and deaths in Cambodia and other similar malaria-endemic regions over the next 10 years. All parameters and initial conditions are given in Supplementary Tables S2 and S3.

We performed a sensitivity analysis on the model, in which each model parameter was modified separately and the relative change in model outputs recorded. To implement the sensitivity analysis, we considered the baseline value of each parameter (given in Tables 2 and 3) and ran simulations with the parameter scaled down to 80\% and up to 120\% while all other parameters remained fixed. If scaling a probability parameter up to 120\% compared to baseline led to a probability being greater than 1 the value was instead held at 1. Similarly, the relative susceptibility of partially-immune individuals compared to susceptible individuals was not scaled up, so as not to exceed 1. For each set of parameter values, 50 repeats of the simulation were conducted and various outputs were recorded, including: the total number of \pf~infections, \pv~infections, clinical \pf~infections, clinical \pv~infections, \pv~relapses, deaths, blood-stage treatments administered and radical cure treatments administered.

\section{Results} \label{sec:results}

\subsection{Scenario modelling}

\begin{figure}
\includegraphics[width=\textwidth]{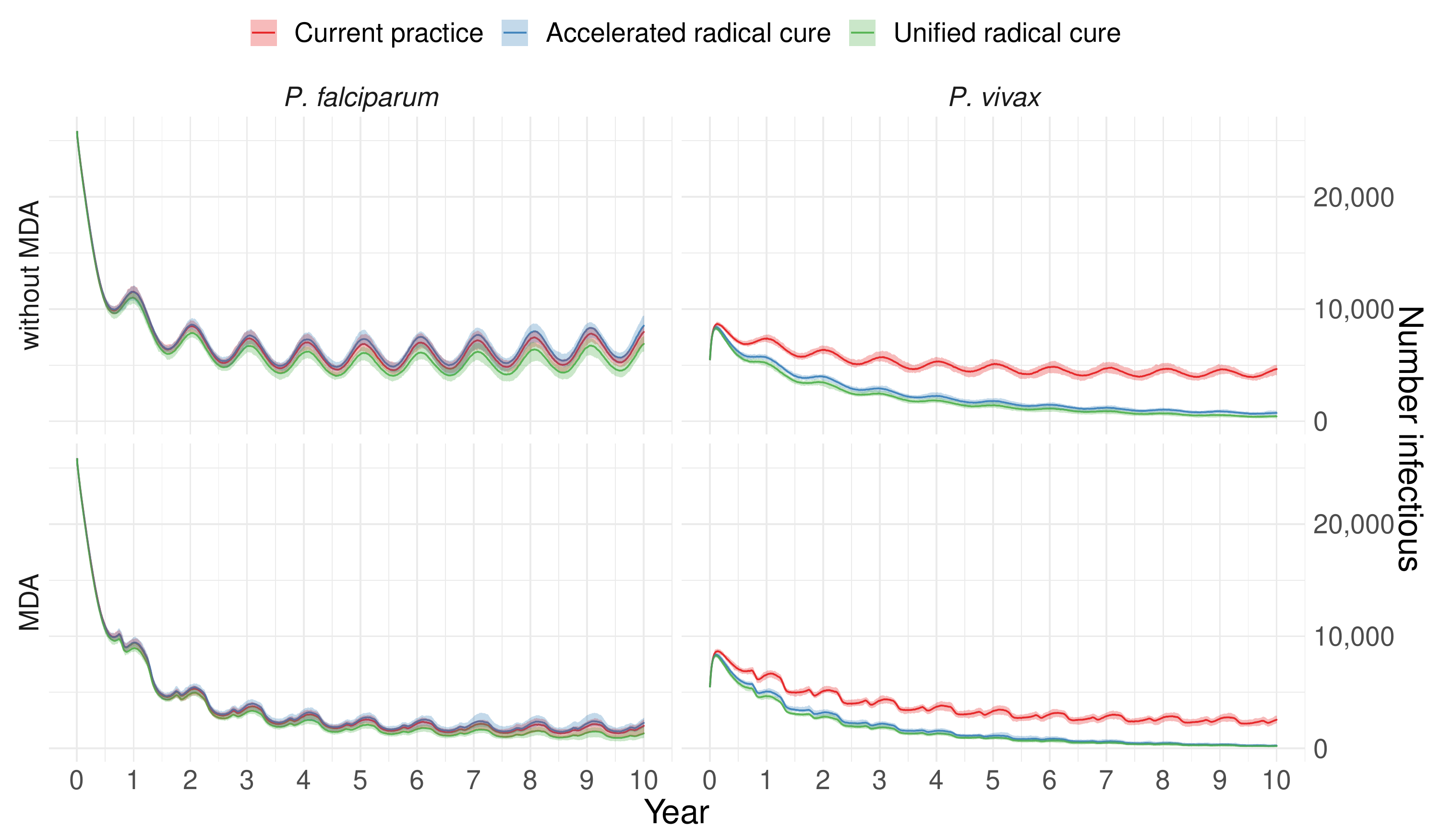}
\caption{Infections over 10 years for \pf~(left panels) and \pv~(right panels) with clinical treatment only (top row), and mass drug administration (MDA) (bottom row).}
\label{fig:time_series}
\end{figure}

Figure \ref{fig:time_series} depicts the total number of infectious individuals in the population over time (the median, minimum and maximum of the 50 simulations) and figures showing all model compartments through time are presented in Supplementary Figures S2 and S3. The different treatment strategies had little effect on the prevalence of \pf, with the unified treatment scenario resulting in a marginally lower prevalence of \pf. For \pv~ increasing the coverage of radical cure has a large impact on the prevalence of \pv, which appears to be approaching elimination towards the end of the time period. Meanwhile, the \ac{mda} intervention greatly reduced the prevalence of \pf~but had less of an effect on \pv. No scenarios led to elimination of malaria over the ten year period by 2030, although elimination may have been achieved over a longer time-frame.

\begin{figure}
\includegraphics[width=\textwidth]{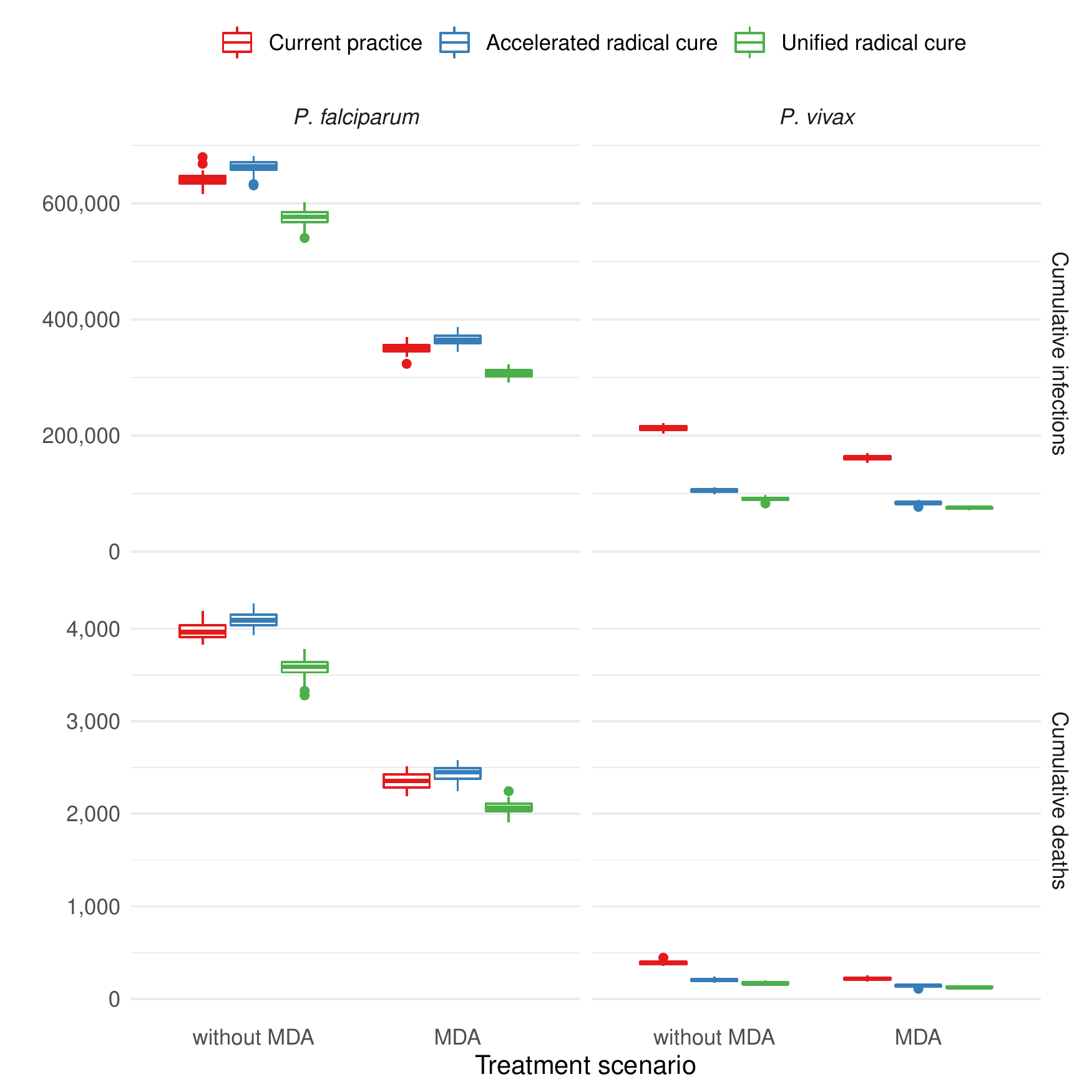}
\caption{Cumulative infections and deaths over a 10 year period with and without mass drug administration (MDA) for \pf~(left panels) and \pv~(right panels).}
\label{fig:box_plots}
\end{figure}

Figure \ref{fig:box_plots} shows boxplots of the cumulative number of infections and deaths by species over the 10 year period. The unified treatment strategy with G6PD testing of all individuals resulted in fewer infections and fewer deaths due to \pf~and \pv, despite the increased risk of haemolysis from radical cure. The accelerated radical cure approach approximately halved \pv~infections but resulted in a small increase in \pf~infections, this increase is likely due to a reduction in the prevalence of mixed infections, which are more likely to be detected than mono-infections, inadvertently causing a reduction in the detection rate of \pf.

\subsection{Sensitivity analysis}

\begin{figure}
\begin{centering}
\includegraphics[width=0.9\textwidth]{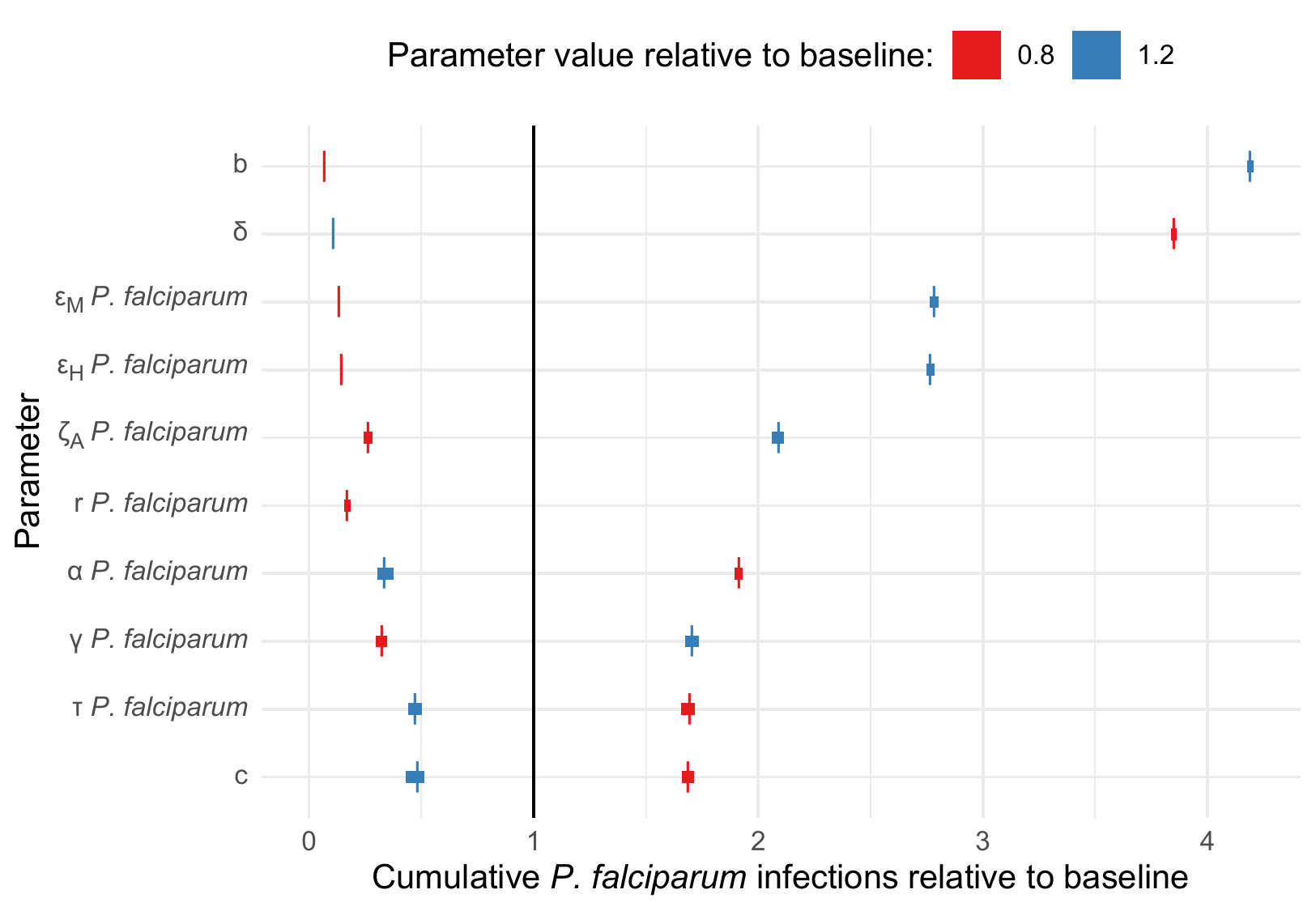}\\
\includegraphics[width=0.9\textwidth, trim = 0 0 0 1.2cm, clip]{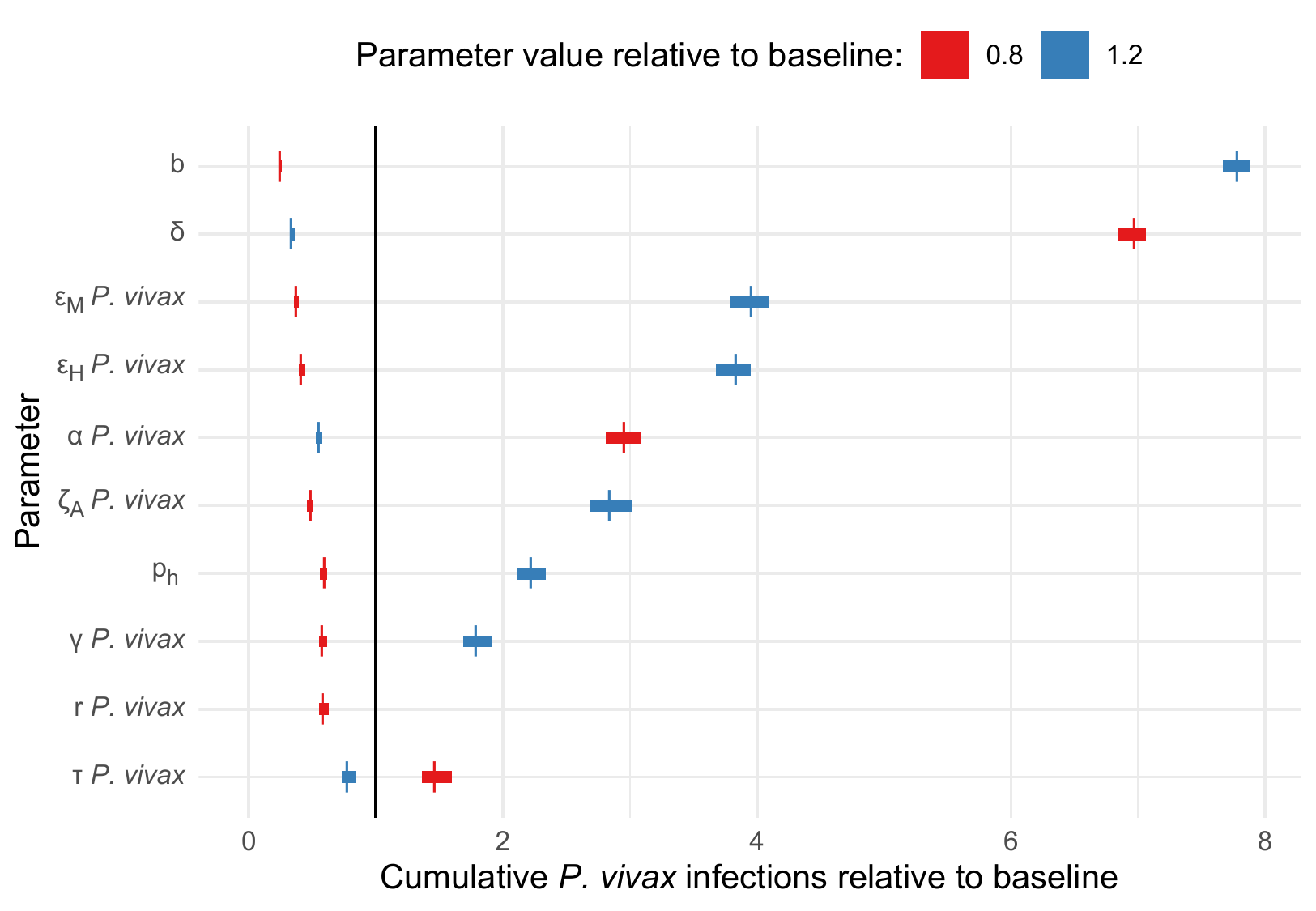}
\caption{Sensitivities of \pf~(top) and \pv~(bottom) cumulative infections with respect to varying model parameters. These are presented in terms of the mean relative outcome, compared to baseline, when each parameter is scaled by 0.8 and 1.2. Red and blue vertical lines represent the mean outcomes relative to baseline given a parameter scaling of 0.8 and 1.2, respectively. Error bars represent the minimum and maximum relative outcome, compared to baseline. Each minimum, mean and maximum calculated from 50 simulations.} 
\label{fig:sensitivity}
\end{centering}
\end{figure}

  In Figure \ref{fig:sensitivity}, the results of the sensitivity analysis are presented in terms of the ten most influential parameters on cumulative infections (including clinical and asymptomatic cases). Sensitivity analyses with respect to all parameters and other outcomes are given in Supplementary Figure S4. These figures present the mean, minimum and maximum relative outcome compared to the baseline over 50 simulations, for each parameter set and orders them based on their relative sensitivity (in terms of absolute difference between the 80\% and 120\% scenarios).

In Figure \ref{fig:sensitivity} we see that \pfv~infections were most sensitive to many of the vector-related parameters, including: the bite rate ($b$), the death rate of mosquitoes ($\delta_0$), the probability of transmission given an infectious bite (from humans to mosquitoes and vice versa, $\epsilon_{M,x}$ and $\epsilon_{H,x}$) and the rate at which exposed mosquitoes become infectious ($\gamma_x$). The spread of mosquito-borne infectious diseases are well known to be sensitive to these parameters \cite{chitnis_determining_2008}.

Aside from the vector-related parameters, we identified several important human-related parameters, including: the relative infectiousness of asymptomatic carriers ($\zeta_{A,x}$), the relative susceptibility of partially-immune individuals ($r$), the rate at which asymptomatic infections are cleared ($\alpha_x$), the rate of treatment seeking ($\tau_x$) and accessibility of treatment ($c$). The parameters $\zeta_{A,x}$ and $\alpha_x$ determine the expected number of secondary infections generated by asymptomatic individuals. The parameter $r$ is related to anti-parasite immunity, it represents a possible lower rate of infection in recovered individuals. The parameters $\tau_x$ and $c$ determine the rate at which clinical cases seek treatment and the probability that they receive treatment for malaria.

In addition to the parameters that were influential on cumulative infections of both species, \pv~infections were sensitive to the probability that hypnozoites are cleared when blood-stage parasites are cleared without treatment ($p_h$). This emphasises the contribution of relapses in the overall \pv~malaria burden.

Other model outcomes (clinical infections, total deaths, total relapses, total treatments received) were broadly sensitive to the same model parameters without any additions or omissions (see Supplement Figure 4 for details). Small differences in relative sensitivity between measuring cumulative infections versus clinical infections were largely centred on the parameters having to do with the proportions immune expected to develop clinical malaria upon reinfection (that is, $p_R$ for \pf~and $p_c$ for \pv).

\section{Discussion} \label{sec:discussion}
Capturing the dynamics of multiple malaria species concurrently is policy-relevant but to date has drawn little attention. In this paper, we have developed a model of sufficient complexity to capture these dynamics, demonstrating how it can be used to inform health policy. Our model incorporates the dynamics of both \pfv~in a way that captures masking, treatment entanglement and triggering interactions of the species. This is the second multi-species model which meaningfully captures dependencies between \pfv~\cite{silal_malaria_2019} and is the only stochastic, agent-based model to do so. The stochasticity in our approach makes it particularly well suited to model \pfv~in low transmission settings, small populations, or as malaria is approaching elimination. The model also has fewer compartments, and fewer parameters, than the only other model with similar features \cite{silal_malaria_2019}, making our model more readily parameterised.

Our scenario analysis explored the effect of different proportions of coverage with radical cure treatment, assuming that individuals were screened for \ac{gpd} deficiency prior to treatment. The scenario analysis showed that a unified radical cure strategy can reduce the prevalence of malaria cases and deaths overall, even when accounting for the increased risk of death due to haemolysis under radical cure. Radical cure is effective because it blocks transmission, kills blood-stage parasites and kills dormant hypnozoites. A unified radical cure strategy avoids issues associated with masking when administering targeted treatment, allows for a consistent protocol for malaria treatment, and does not require the speciation of malaria prior to treatment.

Modelling a twice-yearly \ac{mda} intervention allowed us to assess the additional impact that could be achieved by treating asymptomatic infections as a means to reduce malaria burden. We found that \ac{mda} is an effective way to reduce prevalence, but it will not necessarily lead to elimination if coverage is too low. This is in line with a report from WHO, based on a systematic review of 270 literature reports, which states that for \ac{mda} to be effective, at least 80\% of the population should be treated \cite{who_mda_2015}. Achieving those levels of population coverage may be difficult due to issues with compliance. Due to safety concerns about treating individuals with \ac{gpd} deficiency with radical cure, only \ac{mda} with blood-stage treatment was considered. Consequently we found that \ac{mda} decreased \pf~prevalence but had less of an effect on \pv~prevalence. Targeted interventions may allow radical cure to be administered \textit{en masse} (such as focal screen and treat, or mass screen and treat) \cite{who_mda_2015}. Our modelling framework easily allows these other kinds of interventions to be incorporated through the time-varying treatment function, $\eta_x(t)$.

The sensitivity analysis shows that, although the model has many parameters, the outputs are sensitive to relatively few parameters. The most sensitive parameters for both species were those related to vector-dynamics, the bite rates, transmission probabilities, mosquito death rate and the infectious period of mosquitoes, which are well known to be influential for mosquito-spread diseases \cite{chitnis_determining_2008}. Additionally, the relative infectiousness of asymptomatic individuals ($\zeta_{A,x}$), the rate at which asymptomatic infections are cleared ($\alpha_x$), the relative susceptibility of partially-immune individuals ($r$), the rate of treatment seeking ($\tau$) and accessibility of treatment ($c$) were all found to be influential. Note that $\tau$ and $c$ determine the proportion of clinical infections that go untreated and become asymptomatic. Further, $\zeta_{A,x}$ and $\alpha_x$ determine the expected number of secondary infections generated by asymptomatic individuals. These sensitivities highlight how important asymptomatic individuals can be in driving malaria burden and the need for interventions that target asymptomatic infections, such as \ac{mda} (explored in this work) or better diagnostics that can detect infections with low level parasitaemia. 

The number of \pv~infections was sensitive to the probability of asymptomatic carriers naturally clearing hypnozoites, reinforcing the notion that relapses contribute significantly to malaria burden, as has been shown empirically in an analysis of 68,361 patients in Indonesia \cite{dini_risk_2020}. It is worth noting that since we performed a one-dimensional sensitivity analysis, the results should only be interpreted as the output sensitivity with respect to each parameter in isolation, and not interpreted as a full quantification of model output variation. 
 A full probabilistic sensitivity analysis is appropriate for assessing output uncertainty, particularly if using the model to inform public health policy.

The simulations in our scenario analyses show behaviour comparable to Cambodia with parameters consistent with literature and expert elicitation. Many of the model parameters are location-specific such as the bite rate, relative infectiousness of asymptomatic carriers, probability of death from radical cure and the initial model state. Future work will develop a statistical framework for fitting this model, so that it can be applied in other endemic settings where parameters may differ substantially. The complexity of the multi-species model poses a challenge to jointly fitting all model parameters because of the high dimension of the parameter space and the run time, which was on the scale of minutes for a single simulation over 10 years. Optimised approximate Bayesian inference methods such as Bayesian Optimization for Likelihood-Free Inference (BOLFI) and Likelihood-Free Inference by Ratio Estimation (LFIRE) may provide solutions to both of these challenges \cite{BOLFI_2016,LFIRE_2022}. If the run time of the stochastic model becomes prohibitive for inference, as may be the case when applied to larger populations with high prevalence (where capturing small fluctuations in low numbers is less important), a deterministic or hybrid model equivalent could be applied instead.

 This modelling framework provides the basis for future malaria modelling studies to evaluate the impact of integrated malaria control packages in settings where \pfv~are co-endemic. In particular, parameters in the model can be adjusted to consider vector control measures and to evaluate other treatments, such as single-dose tafenoquine, high-dose 7-day primaquine and triple ACTs. The multi-species malaria model was developed in a way that enables economic analyses through the separation of different treatments and outcomes for individuals given their treatment. In the future, costs and quality of life metrics can be evaluated alongside the impact on cases and deaths. For example, this model could be used to identify under which circumstances a unified treatment for malaria would be cost-effective. Lastly, the modelling framework could be expanded to include other species of malaria, such as zoonotic \textit{P. knowlesi}.

\subsection{Role of the funding source}
ACREME funded the salary of RIH, and contributed to the costs of data cleaning and organisation by PN and the CNM.

\section*{Supplementary Material}
Code and other supplementary material are provided on GitHub at \url{https://github.com/jnwalker2/multispecies-malaria-model}.

\section*{Acknowledgements} \label{sec:acknowledgements}
 This work is supported in part by the Australian Centre for Research Excellence in Malaria Elimination (ACREME), funded by the NHMRC (1134989). J.A. Simpson is funded by an Australian National Health and Medical Research Council of Australia (NHMRC) Investigator Grant (1196068). R. N Price  is a Wellcome Trust Senior Fellow in Clinical Science (200909). J.M. McCaw’s research is supported by the ARC (DP170103076, DP210101920) and ACREME. J.A. Flegg’s research is supported by the ARC (DP200100747, FT210100034). A. Devine's research is supported by National Health and Medical Research Council of Australia (NHMRC) (APP1132975). The contents of the published material are solely the responsibility of the individual authors and do not reflect the views of NHMRC. A. Devine and D.J. Price's research are supported by DFAT.




\bibliographystyle{unsrt}
\bibliography{cospecies-references}






\end{document}